\documentclass[letterpaper,10pt,twocolumn]{article}

\usepackage{usenix}
\usepackage{times}
\PassOptionsToPackage{hyphens}{url}\usepackage[hidelinks]{hyperref}
\usepackage{graphicx}
\usepackage{epstopdf}

\begin{document}

\title{\Large \bf Engineering Record And Replay For Deployability
\\
Extended Technical Report}
\author{
{\rm Robert O'Callahan}\thanks{Majority of work carried out while supported by Mozilla Research.}
\and
{\rm Chris Jones}\footnotemark[1]
\and
{\rm Nathan Froyd} \\
Mozilla Corporation
\and
{\rm Kyle Huey}\footnotemark[1]
\and
{\rm Albert Noll}\footnotemark[1] \\
Swisscom AG
\and
{\rm Nimrod Partush}\footnotemark[1] \\
Technion
}
\date{}
\maketitle
\thispagestyle{empty}

\newcommand{\system}{\textsc{rr}}
\newcommand{\ptrace}{{\tt ptrace}}

\begin{abstract}
The ability to record and replay program executions with low overhead
enables many applications, such as reverse-execution debugging,
debugging of hard-to-reproduce test failures, and ``black box''
forensic analysis of failures in deployed systems. Existing
record-and-replay approaches limit deployability by recording an entire virtual machine
(heavyweight), modifying the OS kernel (adding deployment and maintenance costs),
requiring pervasive code instrumentation (imposing significant performance and complexity overhead), or modifying
compilers and runtime systems (limiting generality). We investigated
whether it is possible to build a practical record-and-replay system avoiding all
these issues. The answer turns out to be yes --- if the CPU and
operating system meet certain
non-obvious constraints. Fortunately modern Intel CPUs, Linux
kernels and user-space frameworks do meet these constraints, although this has only become true recently. With
some novel optimizations, our system \system{} records and replays real-world low-parallelism
workloads with low overhead, with an entirely user-space implementation, using stock hardware, 
compilers, runtimes and operating systems. \system{} forms the basis
of an open-source reverse-execution debugger seeing significant use in practice.
We present the design and implementation of \system{}, describe its performance on a
variety of workloads, and identify constraints on hardware and operating system design required
to support our approach.
\end{abstract}

\section{Introduction}

The ability to record a program execution with low overhead and play it back
precisely has many applications \cite{Devecsery2014,Dolan-Gavitt2015,Engblom2012} and has received significant attention
in the research community. It has even been implemented in products such as VMware Workstation \cite{Malyugin2007}, Simics \cite{Engblom2010}, UndoDB \cite{UndoDB} and TotalView \cite{Gottbrath2008}.
Unfortunately, deployment of these techniques has been limited, for
various reasons. Some approaches \cite{Dunlap2002,Engblom2010,Malyugin2007} require recording and replaying an entire
virtual machine, which is heavyweight --- users must set up and manage the virtual machine, and the state that must be recorded and replayed encompasses an entire operating system, when the user often only cares about some specific process(es). This is especially problematic in record-and-replay applications such as reverse-execution debugging, where checkpoints of the replayed state are frequently created and resumed. Other approaches \cite{Bergan2010,Devecsery2014,Laadan2010,Mashtizadeh2017} require running a modified OS kernel, hindering deployment
and adding security and stability risk to the system. Requiring compiler and language runtime changes \cite{Mashtizadeh2017} also hinders deployment, especially when applications include their own JIT compilers. Some approaches \cite{Hower2008,Montesinos2008,Pokam2013} require custom hardware not yet available. Many approaches \cite{UndoDB,Bhansali2006,Gottbrath2008,Patil2010} require
pervasive instrumentation of code, which adds complexity and overhead, especially for self-modifying code (commonly used in polymorphic inline caching \cite{Holzle1991} and other implementation techniques in modern just-in-time compilers). A performant dynamic code instrumentation engine is also expensive to build and maintain.

Mozilla developers observed that developers spend a lot of time debugging, particularly on
deterministic bugs. Record-and-replay debugging seemed like a promising technology but
none of the existing systems were easy to deploy, for the above reasons. We initiated
the \system{} project to try to build a system suitable for Mozilla's use --- or else discover reasons
why such a system could not be built (with moderate implementation effort).

Therefore we set out to build a system that maximizes deployability: to record and replay
unmodified user-space applications with stock Linux kernels, compilers, language runtimes, and x86/x86-64 CPUs, with a fully user-space implementation running without special
privileges, and without using pervasive code instrumentation. Given our limited development
resources (a couple of person-years for a prototype able to debug Firefox, perhaps five person-years total
to date), we avoid approaches that demand huge engineering effort.
We assume \system{} should run unmodified applications, and they will have bugs (including data races) that we wish to faithfully record
and replay, but these applications will not maliciously try to subvert recording or replay.
We combine techniques already known, but not previously demonstrated working together in a practical
system: primarily, using \ptrace{} to record and replay system call results and signals,
avoiding non-deterministic data races by running only one thread at a time,
and using CPU hardware performance counters to measure application progress so asynchronous signal
and context-switch events are delivered at the right moment \cite{Olszewski2009}.
Section \ref{design} describes our approach in more detail.

With that in place, we discovered the main performance bottleneck for
low-parallelism workloads (in particular, Firefox running the Firefox test suite) was
context switching induced by using \ptrace{} to monitor system calls. We implemented
a novel {\it in-process system-call interception} technique to eliminate those context switches, dramatically
reducing recording and replay overhead on important real-world workloads. This optimization relies on modern Linux kernel
features: {\tt seccomp-bpf} to selectively suppress \ptrace{} traps for certain system
calls, and {\tt perf} context-switch events to detect recorded threads blocking in the kernel.
Section \ref{syscallbuf} describes this work, and Section \ref{results} gives some performance
results, showing that on important application workloads \system{} recording and replay slowdown is less than a factor of two.

We rely on hardware and OS features designed for other goals, so it is surprising that
\system{} works. In fact, it skirts the edge of feasibility; in particular
it cannot be implemented on ARM CPUs. Section \ref{constraints} summarizes \system{}'s hardware and software requirements, which we hope will influence system designers.

\system{} is in daily use by many developers as the foundation of an efficient reverse-execution debugger that works on complex applications such as Samba, Firefox, Chromium, QEMU, LibreOffice and Wine. It is free software, available at \url{https://github.com/mozilla/rr}. This paper makes the following research contributions:
\begin{itemize}
\item We show that record and replay of user-space processes
on modern, stock hardware and software without pervasive code instrumentation is possible and practical.
\item We introduce an {\it in-process system-call interception} technique and show it dramatically reduces overhead.
\item We show that for low-parallelism workloads, \system{} recording and replay overhead is reasonably low, lower than other approaches with comparable deployability.
\item We identify hardware and operating system design constraints required to support our approach.
\end{itemize}

This extended technical report elaborates on published papers with additional technical details and,
in Section \ref{reflection}, a reflection on the design, usage and evolution of \system{}.

\section{Design} \label{design}

\subsection{Summary}

Most low-overhead record-and-replay systems depend on the observation that CPUs are mostly
deterministic. We identify a boundary around state and computation, record all sources
of nondeterminism within the boundary and all inputs crossing into the boundary, and reexecute
the computation within the boundary by replaying the nondeterminism and inputs. If all inputs and
nondeterminism have truly been captured, the state and computation within the boundary during replay
will match that during recording.

To enable record and replay of arbitrary Linux applications, without requiring kernel modifications or
a virtual machine, \system{} records and replays the user-space execution of a group of processes. To simplify
invariants, and to make replay as faithful as possible, replay preserves almost
every detail of user-space execution. In particular, user-space memory and register values are preserved
exactly, with a few exceptions noted later in the paper. This implies CPU-level control flow is
identical between recording and replay, as is memory layout.

While replay preserves user-space state and execution, only a minimal amount of kernel state is
reproduced during replay. For example, file descriptors are not opened, signal handlers are not
installed, and filesystem operations are not performed. Instead the recorded user-space-visible effects of those
operations, and future related operations, are replayed. We do create one replay thread per
recorded thread (not strictly necessary), and we create one replay address
space (i.e.\ process) per recorded address space, along with matching memory mappings.

With this design, our recording boundary is the interface between user-space and the kernel.
The inputs and sources of nondeterminism are mainly the results of system calls, and the timing of
asynchronous events.

\subsection{Avoiding Data Races}

With threads running on multiple cores, racing read-write or write-write accesses to
the same memory location by different threads would be a source of
nondeterminism. Therefore we take the common approach \cite{Dunlap2002, Malyugin2007, UndoDB, Dolan-Gavitt2015} running only one
thread at a time. \system{} preemptively schedules these threads, so context switch timing is
 nondeterminism that must be recorded. Data race bugs can still be observed if a context
switch occurs at the right point in the execution (though bugs due to weak memory models cannot be
observed).

This approach is much simpler and more deployable than alternatives \cite{Bhansali2006, Dunlap2008, Patil2010, Veeraraghavan2011,
Mashtizadeh2017}, avoids assuming programs are race-free \cite{Devecsery2014, Mashtizadeh2017}, and is efficient for low-parallelism workloads. 
There is a large slowdown for workloads with a consistently high degree of
parallelism; however, even for applications which are potentially highly parallel,
users often apply \system{} to test workloads with relatively small
datasets and hence limited parallelism.

The one-thread-at-a-time restriction is implemented by supervising all tracee processes/threads via
the \ptrace{} system call. Whenever a thread enters a system call,
\system{} receives a ``ptrace stop'' event from the kernel and decides whether to allow a context
switch or continue running the current thread. (At any potentially blocking system call, we must
allow a context switch.) When a context switch is allowed, \system{} uses {\tt PTRACE\_SYSCALL}
to let the thread continue running to system call exit, then selects another thread to run and resumes it with
a {\tt PTRACE\_CONT} (or similar) \ptrace{} operation. When the new thread in turn reaches a reschedule
point, \system{} queries the
state of all threads using {\tt waitpid} to determine which, if any, are ready to exit blocking
system calls and are therefore candidates for scheduling.

\system{} performs preemptive context switching by periodically interrupting the running thread
with an asynchronous signal (see Section \ref{async-events}). The normal scheduler honors Linux thread
priorities strictly: higher-priority threads always run in preference to lower-priority threads. Threads
with equal priority run in a round-robin manner.

\subsection{System Calls} \label{syscalls}

System calls return data to user-space by modifying registers and memory, and these changes must be recorded. The \ptrace{}
system call allows a process to be synchronously notified when a tracee thread enters or exits a system call. 
When a tracee thread enters the kernel for a system call, it is suspended and
\system{} is notified. When \system{} chooses to run that thread again, the system call will complete, notifying
\system{} again, giving it a chance to record the system call results. \system{} contains a model of most Linux system calls
describing the user-space memory they can modify, given the system call input parameters and result.

\subsubsection{Scratch Buffers}

As noted above, \system{} normally avoids races by scheduling only one thread at a time. However,
if a system call blocks in the kernel, \system{} must try to schedule another application thread to run
while the blocking system call completes. It's possible (albeit unlikely) that the running thread
could access the system call's output buffer and race with the kernel's writes to that buffer. To avoid this, we redirect
system call output buffers to per-thread temporary ``scratch memory'' which is otherwise unused by the application.
When we get a \ptrace{} event for a blocked system call completing, \system{} copies scratch buffer
contents to the real user-space destination(s) while no other threads are running, eliminating the race. (We actually have no evidence that the races prevented by scratch buffers occur in practice, and it might be worth trying to eliminate scratch buffers.)

\begin{figure}[t]
\centering
\includegraphics[scale=0.4]{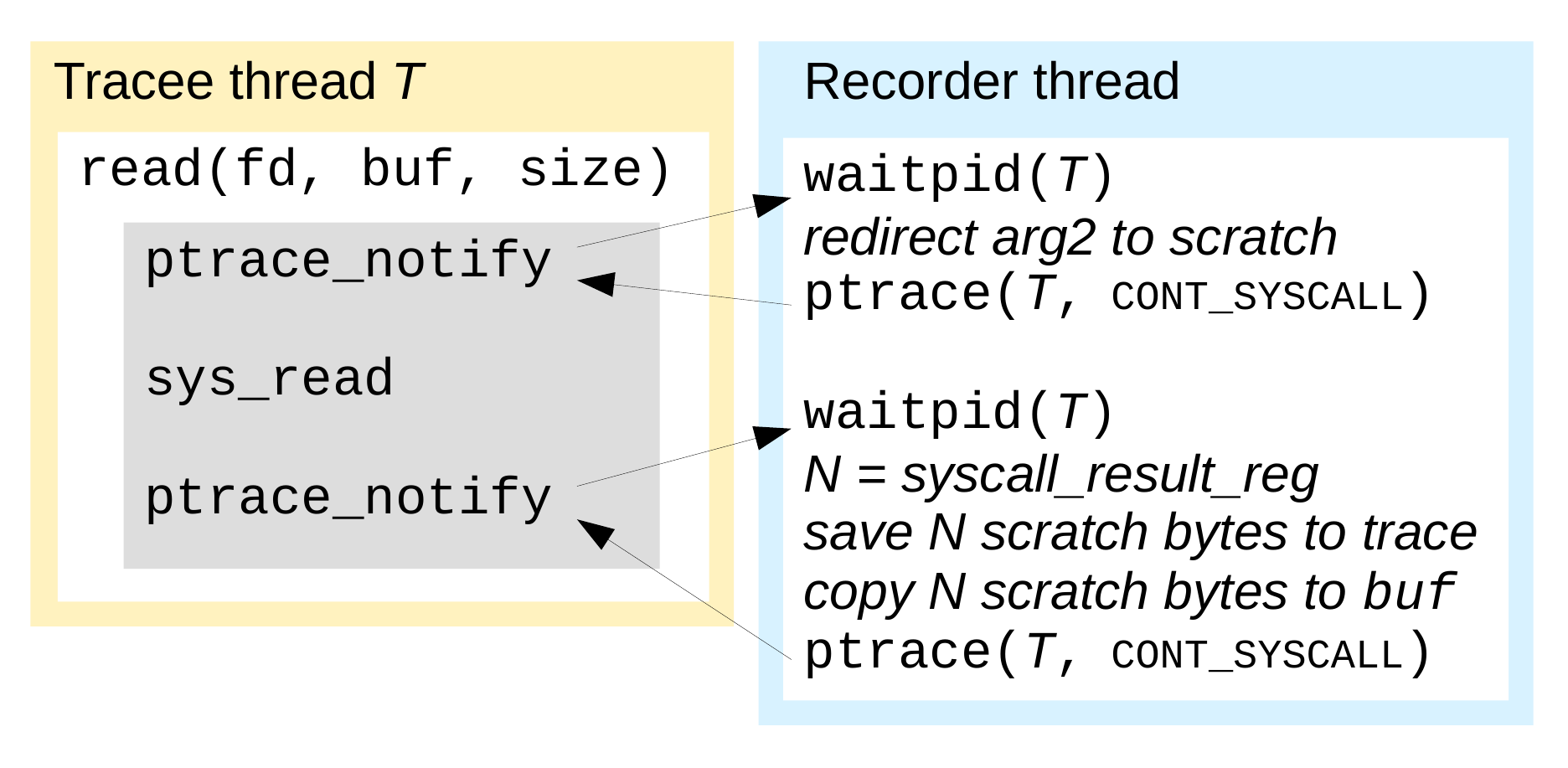}
\caption{Recording simple system call}
\label{simple-syscall}
\end{figure}

Figure \ref{simple-syscall} illustrates recording a simple {\tt read} system call. The gray box represents kernel code.

\subsubsection{\ptrace{} Emulation}

Apart from redirection through scratch buffers and recording memory effects, most system calls execute normally during recording.
One important exception is \ptrace{}. \ptrace{} is not only used by debuggers but also by some
sophisticated applications, e.g. to collect data for crash reporting in Firefox, or to provide emulation
facilities in Wine \cite{Wine}. Linux only allows a thread to have a single
\ptrace{} supervisor, so if one tracee
tried to \ptrace{} another directly, that would fail because the latter already has \system{} as a
\ptrace{} supervisor. Instead \system{} emulates all tracee
\ptrace{} operations and relationships. \ptrace{} semantics are complicated, so this
\ptrace{} emulation is also necessarily rather complicated to implement.

\subsubsection{Performing System Calls In Tracee Context}

In a variety of situations (e.g. {\tt mmap} handling, see below)
\system{} must perform arbitrary system calls in the context of a tracee
thread. \system{} locates a system-call instruction in the tracee's address space and uses \ptrace{}
to temporarily alter the tracee execution context to execute that instruction with the desired parameters in
registers. Sometimes memory parameters are also required, in which case readable memory is located on the
stack (or \emph{in extremis} anywhere in the address space), temporarily overwritten with the desired
parameters, and then reset to its original contents after the system call has completed.

\subsubsection{Kernel Namespaces}

Linux kernel namespaces are a powerful feature used to implement ``containers'', and are also used
to implement sandboxing in sophisticated applications such as Firefox and Chromium. They mostly
require no special handling, except that it is sometimes necessary for a sandboxed process to
communicate with \system{}, e.g. by opening shared files or sockets. To make this possible,
we open a hole in the sandbox by allocating a reserved file
descriptor in all tracees that refers to the root directory as seen by \system{}; with this
descriptor, a tracee can open temporary files or sockets created by \system{} even if its mount namespace
or root directory have been changed.

A malicious but sandboxed tracee aware of this hole could exploit it to escape from the
sandbox, which is one reason we assume tracees are not malicious.

\subsubsection{Seccomp}

Another Linux sandboxing feature is {\tt seccomp} filtering. This installs a {\tt seccomp-bpf} filter
into the kernel, bytecode which evaluates the register parameters of each system call
and determines whether the system call can proceed normally or should produce an error, signal or
{\tt ptrace} trap. Sandboxes typically use this feature to whitelist permitted system calls and
trigger emulation of others. This sandboxing can disrupt \system{} by
interfering with the system calls \system{} performs in the tracee context.

Immediately after each {\tt execve} we map a page of memory at a fixed address, the ``\system{} page'', containing a
``privileged'' system-call instruction. We detect the system calls that install
{\tt seccomp-bpf} filters, and patch the filter bytecode with a prologue exiting
early with ``allow'' if the program counter is at the ``privileged'' instruction address.
We use that ``privileged'' instruction when performing \system{} system calls in tracee context.

We also record the effects of {\tt seccomp-bpf} filters on regular system calls and replay
them faithfully. We do this just by observing kernel behavior, without having to
actually implement a BPF interpreter in \system{}.

\subsubsection{System Call Modeling In Practice} \label{syscall-modeling}

\system{} assumes it is possible to efficiently determine from user-space state before
and after a system call (plus, possibly, observations of previous system calls) reasonable
bounds on the memory locations modified by the system call. For example, we would not
be able to handle a system call that writes to random locations in memory without somehow
indicating which locations were modified. This assumption holds adequately in practice,
but it is a lot of work to create and maintain the model that describes how to determine the memory
modified by each system call.

The biggest problem is the {\tt ioctl} system call, which provides per-device (therefore
extensible) system call namespaces. There is a convention for encoding the address and length of
written memory in the ioctl parameters, but it is often not followed.
\system{} builds in knowledge of many ioctls, but this is necessarily incomplete and needs to
be extended from time to time to handle new applications.

This issue is not unique to \system{} --- for example, Valgrind Memcheck also needs this information.
It would be useful to have a shared repository of system call descriptions from which our
models could be generated.

An alternative approach to modeling all system calls would be to use the Linux "soft-dirty"
tracking API \cite{SoftDirty} to identify pages modified by infrequently used or unknown
system calls. We haven't tried this because although it might ultimately reduce work, it
adds complexity and would not be compatible with applications that use soft-dirty for their
own needs. Also, requiring \system{} to be aware of the semantics of all system calls means that
we detect new system calls that require special handling by \system{}.

\subsubsection{Replay}

During replay, when the next event to be replayed is an intercepted system call, we
set a temporary breakpoint at the address of the system call instruction (recorded in the trace). We use
\ptrace{} to run the tracee thread until it hits
the breakpoint, remove the breakpoint, advance the program counter past the system call instruction,
and apply the recorded register and memory changes. This approach is simple and more efficient
than using {\tt PTRACE\_SYSEMU} to enter and suppress execution of system calls, because it
reduces the number of \ptrace{}-stop notifications from two (on entry to and exit from the system call) to one.
(These notifications are relatively expensive because each one requires context switching from the
tracee to \system{} and back.)

In some cases this technique may not work reliably. For example, the tracee may write a system-call instruction
to memory and then jump to it, in which case a software breakpoint set at that location would be
overwritten. We could use hardware breakpoints but that could conflict with a tracee's own use of
hardware breakpoints, since x86 CPUs support a limited number of them (typically four). Therefore
if the system call instruction to replay is in shared memory or writeable memory, we use
{\tt PTRACE\_SYSEMU} to advance to the execution of the next system call and incur the cost of
two \ptrace{}-stop notifications.

\subsubsection{Replaying Complex System Calls}

Some system calls manipulate threads or address
spaces and require special handling during replay. For example a recorded file {\tt mmap} saves
the original contents of the mapped file to the trace (often in an optimized zero-copy way; see Section
\ref{trace-sizes}). Replay maps that copy with the {\tt MAP\_FIXED} option to ensure the mapping
is created at the correct address.

The {\tt execve} system call is complicated to record and replay. After a recorded {\tt execve} has completed,
\system{} reads {\tt /proc/}...{\tt /maps} to discover the memory mappings in the new address space
and record them. In replay, \system{} {\tt execve}'s a small stub executable for the correct architecture
(32-bit or 64-bit), removes all the memory mappings in the address space (since they may be completely
different from those created during recording, e.g. due to ASLR), then creates new mappings to match
those recorded.

\subsubsection{Signal Handlers}

When a synchronous or asynchronous signal triggers a signal handler during recording, the kernel sets up
a user-space stack frame for the handler and transfers control to the address of the
handler. \ptrace{} reports the delivery of a signal as via a \ptrace{}-stop notification, then
\system{} triggers the transition to the signal handler by issuing a {\tt PTRACE\_SINGLESTEP}
request. (\system{} must track internally which signals have handlers so it knows in advance which
signals will trigger a handler. It also needs to be aware of quirky edge cases that suppress entry into
a signal handler; for example, when code triggerings a {\tt SIGILL} or {\tt SIGSEGV}, but the
tracee has used {\tt sigprocmask} to block the signal, the signal is delivered but handlers do
not run (usually with fatal results).)

When \system{} has transitioned into the signal handler, it records the contents of the stack frame
and registers set up by the kernel. During replay, no signal handlers are set up and no real signals
are delivered.
To replay the entry to a signal handler, we just write the recorded signal handler frame into memory
and set the registers to the recorded values. By not delivering real signals during replay, we avoid
having to set up signal-handler state in the replaying process.

\subsubsection{Signal Handling And Interrupted System Calls}

During normal non-recorded execution, when a system call is interrupted by a signal,
the kernel adjusts the user-space registers with the program counter immediately
after the system-call instruction and a special {\tt -ERESTARTSYS} (or similar) value in the
system-call result register. If a signal handler needs to run, the signal frame is built and
the signal handler runs normally. When execution returns to the pre-handler state via {\tt sigreturn},
or a signal handler did not run but the thread is resumed by some other mechanism, then just before
returning to user-space the
kernel's system-call restart mechanism kicks in: when the kernel returns to userspace, if the
system-call result register contains {\tt -ERESTARTSYS} and certain other conditions are met,
the kernel automatically backs up the program counter to before the system call instruction and
restores the system-call-number register. This causes the system call to reexecute.

\system{} avoids interfering with this restart mechanism and relies on the kernel restarting
interrupted system calls as it normally would. However, because we set up state on system-call
entry that expects to be processed on system-call exit, we have to detect
when a \ptrace{} system-call notification corresponds to a restarted system call instead of
a new system call. There is no foolproof way to do this, but we have a heuristic that works well.
We maintain a stack of pending interrupted system calls and their saved parameter registers.
On every system call entry, if the parameter registers match the parameters of the most recent
pending interrupted system call, we assume that system call is being restarted and pop it off
the stack. If a signal
handler interrupted a system call and then returns, but the interrupted system call is not
immediately restarted, we assume it has been abandoned and pop it off the stack.

\subsection{Asynchronous Events} \label{async-events}

We need to support two kinds of asynchronous events: preemptive context switches and signals.
We treat the former as a special case of the latter, forcing a context switch by sending a signal
to a running tracee thread. We need to ensure that during replay, a signal is delivered when the
program is in exactly the same state as it was when the signal was delivered during recording.

As in previous work \cite{Dunlap2002, Olszewski2009, Burtsev2016} we measure application progress using CPU hardware performance counters. Ideally we would count
retired instructions leading up to an asynchronous event during recording, and during replay program
the CPU to fire an interrupt after that many instructions have been retired --- but this approach needs
modifications to work in practice.

\subsubsection{Nondeterministic Performance Counters} \label{counters}

We require that every execution of a given sequence of user-space instructions changes the counter value by an amount that depends only on the instruction sequence, not system state invisible to user space (e.g.\ the contents of
caches, the state of page tables, or speculative CPU state). (Such effects manifest as noise making
the counter unreliable from \system{}'s point of view.)
This property (commonly described as ``determinism'' \cite{Weaver2013}) does not hold for
most CPU performance counters in practice \cite{Dunlap2002,Weaver2013}. For example it does not hold for any ``instructions retired''
counter on any known x86 CPU model (e.g.\ because an instruction triggering a page fault is restarted and counted twice).

Fortunately, modern Intel CPUs have exactly one deterministic performance counter: ``retired conditional branches'' (``RCB''), so
we use that. We cannot just count the number of RCBs during recording and deliver the signal after we have
executed that number of RCBs during replay, because the RCB count does not uniquely determine the execution point to
deliver the signal at. Therefore we pair the RCB count with the complete state of general-purpose registers (including the program counter)
to identify an execution point.

In general that \emph{still} does not uniquely identify an
execution point (e.g.\ consider the infinite loop {\tt label: inc [global\_var]; jmp label;}).
However, in practice we have found it works reliably; code that returns to the same instruction with no intervening
conditional branch must be very rare, and it only matters to \system{} for regular replay
if an asynchronous event occurs at such an instruction --- in which case replay would probably diverge and fail.

When \system{} is used for more than just plain record-and-replay, in particular to implement reverse-execution
debugging, it becomes considerably more sensitive to situations where the general-purpose registers and the RCB count
do not uniquely identify a program state, because our reverse-execution implementation sometimes
performs very frequent stops (e.g. while single-stepping) and many comparisons of one stopping point to
another. We encountered a few practical
examples where distinct program states with the same general-purpose registers and RCB count would cause
reverse-execution debugging to fail. These were cases where a function A contained multiple calls to the
same function B, no conditional branches occurred in A or B,
and due to optimization, at some instruction in B general-purpose registers ended up with the same values in
both invocations. We addressed this issue by collecting some data from the stack at each stopping point
and adding it to the program-state identifier. This now seems to be robust;
we have not encountered any new issues in this area for a couple of years.

\subsubsection{Alternative Approaches to Nondeterministic Counters}

Other work \cite{Dunlap2002, Burtsev2016} avoids those uniqueness issues by counting the
number of branches retired; an instruction at a given address can't be reached more than once
without an intervening branch (except for repeating string instructions, which require the
{\tt ECX}/{\tt RCX} register to be added to the program-state identifier). Unfortunately
the branch counter is nondeterministic; for example a transition out a hardware interrupt is
counted against the user process, but invisible to \system{}. Even worse, transitions out
of ``Systems Management Mode'' interrupts are counted against the user process and aren't even visible
to the host kernel. Working around these issues at user-level seems harder than the
approach \system{} has been using.

Another interesting issue is that cloud virtualization services have been reluctant to
enable virtualization of hardware performance counters, partly because counter nondeterminism
due to low-level CPU state often represents information leakage from one guest VM to another.
By definition, a deterministic counter is unaffected by information not already visible in
its process, and we hope therefore more likely to be safely virtualizable. (At time of writing,
most cloud providers do not virtualize any performance counters and thus \system{} does not work
on them, except for Digital Ocean, where \system{} does work.)

The other obvious approach to measuring progress is to eschew hardware performance
counters and use binary instrumentation to count instructions instead \cite{Patil2010, Bhansali2006, UndoDB}.
We believe the overhead and implementation complexity of binary instrumentation
still make it less desirable than using hardware counters, for our purposes.

\subsubsection{Late Interrupt Firing}

Another problem with hardware performance counters is that, although CPUs can be programmed to fire an interrupt after a specified number of
performance events have been observed, the interrupt does not fire immediately. In practice we often observe it firing after
dozens more instructions have retired. To compensate for this, during replay, we program the interrupt to trigger
some number of events earlier than the actual RCB count we are expecting. Then we set a temporary breakpoint
at the program counter value for the state we're trying to reach, and repeatedly run to the breakpoint until the RCB count and the general-purpose register values match their recorded values.

\subsection{Shared Memory} \label{shmem}

By scheduling only one thread at a time, \system{} avoids issues with races on shared memory as long as
that memory is written only by tracee threads. It is possible for recorded processes to share memory
with other processes, and even kernel device drivers, where that non-recorded code can perform writes
that race with accesses by tracee threads. Fortunately, this is rare for applications running in common Linux desktop environments, occurring in only four common cases: applications sharing memory with the PulseAudio daemon,
applications sharing memory with the X server, applications sharing memory with kernel graphics drivers and GPUs, and {\tt vdso} syscalls. We avoid the first three problems by
automatically disabling use of shared memory with PulseAudio and X (falling back to a socket transport in both cases), and disabling direct access to the GPU from applications.

{\tt vdso} syscalls are a Linux optimization that implements some common read-only system calls (e.g.\ {\tt gettimeofday}) entirely in user space, partly by reading memory shared with the kernel and
updated asynchronously by the kernel. We disable {\tt vdso} syscalls by patching their user-space implementations
to perform the equivalent real system call instead.

Applications could still share memory with non-recorded processes in problematic ways, though this is rare in practice and
can often be solved just by enlarging the scope of the group of processes recorded by \system{}.

\subsection{Nondeterministic Instructions} \label{instructions}

Almost all CPU instructions are deterministic, but some are not. One common nondeterministic x86 instruction
is {\tt RDTSC}, which reads a time-stamp counter. This particular instruction is easy to handle, since the
CPU can be configured to trap on an {\tt RDTSC} and Linux exposes this via a {\tt prctl} API, so we can
trap, emulate and record each {\tt RDTSC}.

{\tt RDRAND} generates random numbers and hopefully is
not deterministic. We have only encountered it being used in one place in GNU {\tt libstdc++}, so \system{} patches that explicitly.

The {\tt CPUID} instruction is mostly deterministic, but one of its features returns the index of the
running core, which affects behavior deep in {\tt glibc} and can change as the kernel
migrates a process between cores. We use the Linux {\tt sched\_setaffinity} API to force all tracee threads to run
on a particular fixed core, and also force them to run on that core during replay.

We could easily avoid most of these issues in well-behaved programs if we could just trap-and-emulate the {\tt CPUID}
instruction from user-space, since then we could mask off the feature bits indicating support for {\tt RDRAND}, hardware
transactions, etc, as well as record-and-replay the exact values returned by the recording CPU.
Modern Intel CPUs support this (``{\tt CPUID} faulting''); we are in the process of adding a user-accessible API for this to Linux. (In fact it just landed in Linux 4.12.)

\subsubsection{Transactional Instructions}

Modern Intel CPUs support two forms of transactional extensions. ``Restricted Transactional Memory'' (``RTM'') exposes transactions as an architectural feature, using the {\tt XBEGIN} and {\tt XEND} instructions to explicitly start and end transactions. ``Hardware Lock Elision'' (``HLE'') supports implicit transactions by adding the {\tt XACQUIRE} and {\tt XRELEASE} prefixes to the instructions that acquire and release a lock. The latter is a pure backwards-compatible performance hint that does not change visible semantics --- failed transactions are automatically retried non-speculatively. The former has visible side effects due to explicitly exposing transaction failures.

RTM is nondeterministic from the point of view of user-space, since a hardware transaction can succeed or fail
depending on CPU cache state (and probably the occurrence of hardware interrupts). Fortunately so far we have only
found these being used by the system {\tt pthreads} library, and we dynamically apply custom patches to that
library to disable use of hardware transactions. In the future we will use mask off the RTM feature bit in {\tt CPUID} to stop applications using RTM.

HLE does not have user-space-visible side effects, but the regular RCB counter counts conditional branches
retired in a failed HLE transaction, making that counter nondeterministic when HLE is used. We can't
disable HLE via {\tt CPUID}, because HLE being backwards-compatible means code is free to use the HLE prefixes without
checking {\tt CPUID} first. Intel provides an option ({\tt IN\_TXCP}) to have hardware performance counters only count events
in committed transactions, so we can use that to make the RCB counter deterministic again. Unfortunately on Linux
an {\tt IN\_TXCP} counter can't be configured to interrupt after a certain number of events have been observed, because
this could cause infinite loops in some cases. (The scenario is: counter $C$ is configured to interrupt after $N$ events,
a transaction starts and executes $N$ events, the interrupt fires and causes the transaction to be rolled back, restoring the counter to its original value, the kernel resumes the process, repeat...) Our solution is to use a non-{\tt IN\_TXCP} counter to trigger interrupts and a {\tt IN\_TXCP} counter to count elapsed events. This means an interrupt could fire early, which is not a problem since we already program it to fire early.

\subsection{Reducing Trace Sizes} \label{trace-sizes}

For many applications the bulk of their input is memory-mapped files, mainly executable code. Copying
all executables and libraries to the recorded trace on every execution would impose significant time and space overhead.
\system{} creates hard links to memory-mapped executable files instead of copying them; as long as a system update
or recompile replaces executables with new files, instead of writing to the existing files, the links retain the
old file data. This works well in practice.

Even better, modern filesystems such as XFS and Btrfs offer copy-on-write logical copies of files (and even block ranges within files), ideal for our purposes. When a mapped file is on the same filesystem as the recorded trace, and the filesystem supports cloning, \system{} clones mapped files into the trace. These clone operations are essentially free in time and space, until/unless the original file is modified or deleted.

\system{} compresses all trace data, other than cloned files and blocks, with the {\tt zlib} ``deflate'' method. Significantly better compression algorithms exist, and at some point we should revisit this choice.

Nevertheless, with these optimizations, in practice trace storage is a non-issue. Section \ref{space-results} presents some results.

\section{In-process System-call Interception} \label{syscallbuf}

The approach described in the previous section works, but overhead is disappointingly high (see Figure \ref{optimizations-chart} below). The core problem is that for every tracee system call,
as shown in Figure \ref{simple-syscall} the tracee performs four context switches: two blocking \ptrace{} notifications, each requiring a context switch from the tracee to \system{} and back. For common system calls such as {\tt gettimeofday} or {\tt read} from cached files,
the cost of even a single context switch dwarfs the cost of the system call itself. To significantly reduce overhead, we must
avoid context-switches to \system{} when processing these common system calls.

Therefore, we inject into the recorded process a library that intercepts common system calls, performs the system call without triggering
a \ptrace{} trap, and records the results to a dedicated buffer shared with \system{}. \system{} flushes the buffer to its
trace every time it receives a synchronous stop notification. The concept is simple but there are problems to overcome.

\subsection{Intercepting System Calls}

A common technique for intercepting system calls in-process is to use dynamic linking to interpose wrapper functions over the C library functions that make system calls. In practice, we have found that method to be insufficient, due to applications making direct system calls, and fragile, due to variations in C libraries, and applications that require their own preloading \cite{Serebryany2012, Wine}).

Instead, when the tracee makes a system call, \system{} is notified via a \ptrace{} trap and it tries to
rewrite the system-call instruction to call into our interception library. This is tricky
because on x86 a system call instruction is two bytes long, but we need to replace it with a five-byte {\tt call}
instruction.
(On x86-64, to ensure we can call from anywhere in the address space to the interception
library, we also need to also allocate trampolines within 2GB of the patched code.)
In practice, frequently executed system call instructions are followed by a few known, fixed instruction sequences;
for example, many system call instructions are followed by a {\tt cmpl \$0xfffff001,\%eax} instruction testing
the syscall result. We added five hand-written stubs to our interception library that
execute post-system-call instructions before returning to the patched code. On receipt of a \ptrace{} system-call notification, \system{}
replaces the system call instruction and its following instruction with a call to the corresponding stub.

We (try to) redirect all system call instructions to the interception library,
but for simplicity it only contains wrappers for the most common system calls, and for
others it falls back to doing a regular \ptrace{}-trapping system call.

\subsection{Selectively Trapping System Calls}

\ptrace{} system-call monitoring triggers traps for all system calls, but our
interception library needs to avoid traps for selected system calls. Fortunately,
modern Linux kernels support selectively generating \ptrace{} traps: {\tt seccomp-bpf}.
{\tt seccomp-bpf} was designed primarily for sandboxing. A process can apply a {\tt seccomp-bpf} filter function, expressed in bytecode, to another process; then, for every system call performed by the target process, the kernel runs the filter, passing in incoming user-space register values, including the program counter. The filter's result directs the kernel to either allow the system call, fail with a given {\tt errno}, kill the target process, or trigger a \ptrace{} trap. Overhead of filter execution is
negligible since filters run directly in the kernel and are compiled to native code on most architectures.

\begin{figure}[t]
\centering
\includegraphics[scale=0.4]{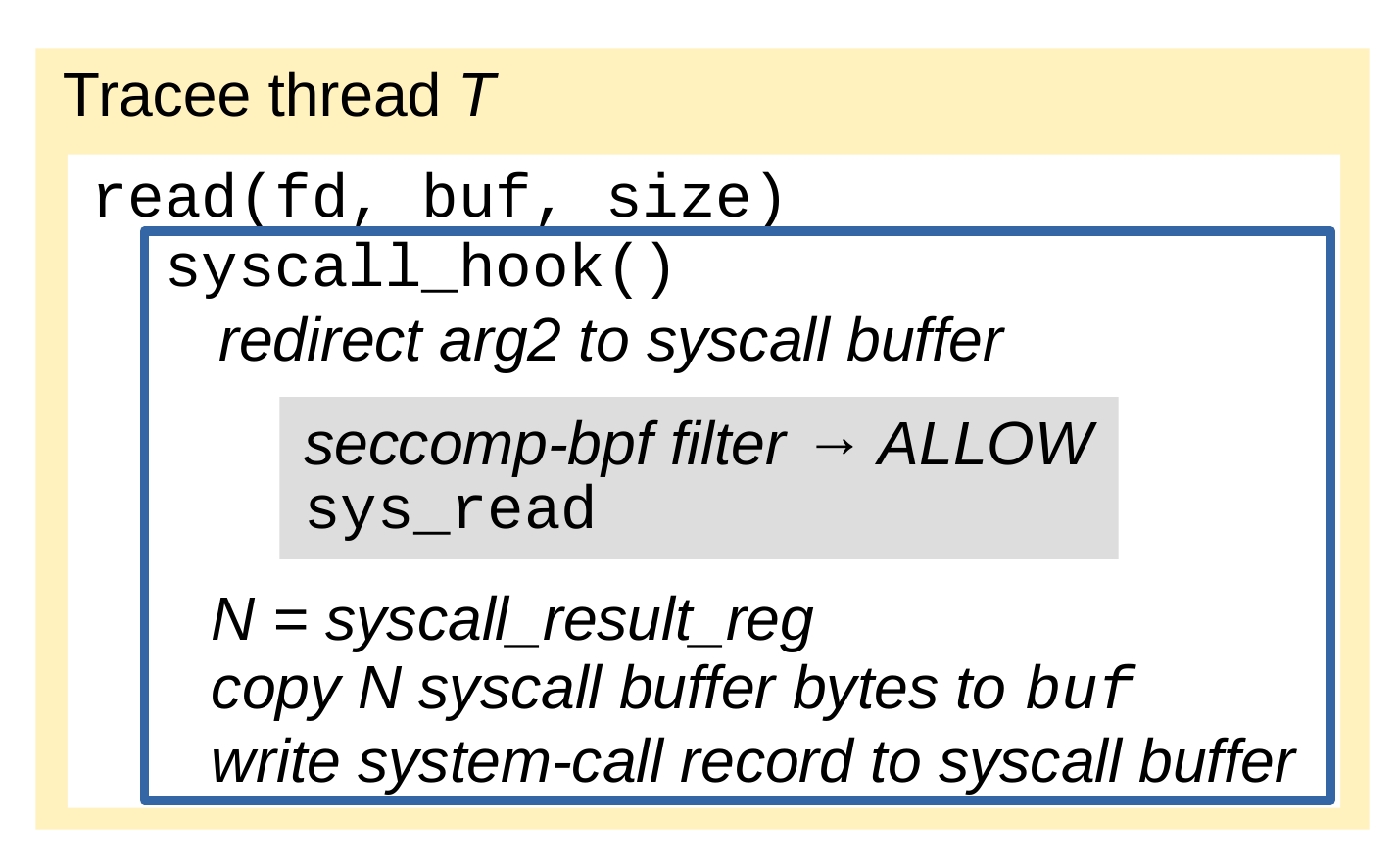}
\caption{Recording with system-call interception}
\label{syscallbuf-diagram}
\end{figure}

Figure \ref{syscallbuf-diagram} illustrates recording a simple {\tt read} system call with in-process system-call interception. The solid-border box represents code in the interception library and the grey box represents kernel code.

As mentioned above, \system{} injects a ``\system{} page'' into every tracee process at a fixed address.
That page contains a special system call instruction --- the ``untraced instruction''.
\system{} applies a {\tt seccomp-bpf} filter to each recorded process that triggers a \ptrace{} trap for
every system call --- except when the program counter is at the untraced instruction, in which case
the call is allowed. Whenever the interception library needs to make an untraced system call, it uses that
instruction.

\subsection{Detecting Blocked System Calls}

Some common system calls sometimes block (e.g.\ {\tt read} on an empty pipe). Because \system{} runs
tracee threads one at a time, if a thread enters an untraced blocking system call without notifying \system{}, it
will hang and could cause the entire recording to deadlock (e.g.\ if another tracee thread is about to
{\tt write} to the pipe). We need the kernel to notify \system{} and suspend the tracee thread whenever
an untraced system call blocks, to ensure we can schedule a different tracee thread.

We do this using the Linux {\tt perf} event system to monitor {\tt PERF\_COUNT\_SW\_CONTEXT\_SWITCHES}.
The kernel raises one of these events every time it deschedules a thread from a CPU core. 
The interception library monitors these events for each thread and requests that the
kernel send a signal to the blocked thread every time the event occurs. These signals trigger \ptrace{}
notifications to \system{} while preventing the thread from executing further.
To avoid spurious signals (e.g.\ when the thread is descheduled due to normal timeslice expiration), the event
is normally disabled and explicitly enabled during an untraced system call that might block. Still, spurious {\tt SWITCHES} can occur at any point between enabling and disabling the event; we handle these edge cases with careful inspection of the tracee state.

\begin{figure}[t]
\centering
\includegraphics[scale=0.4]{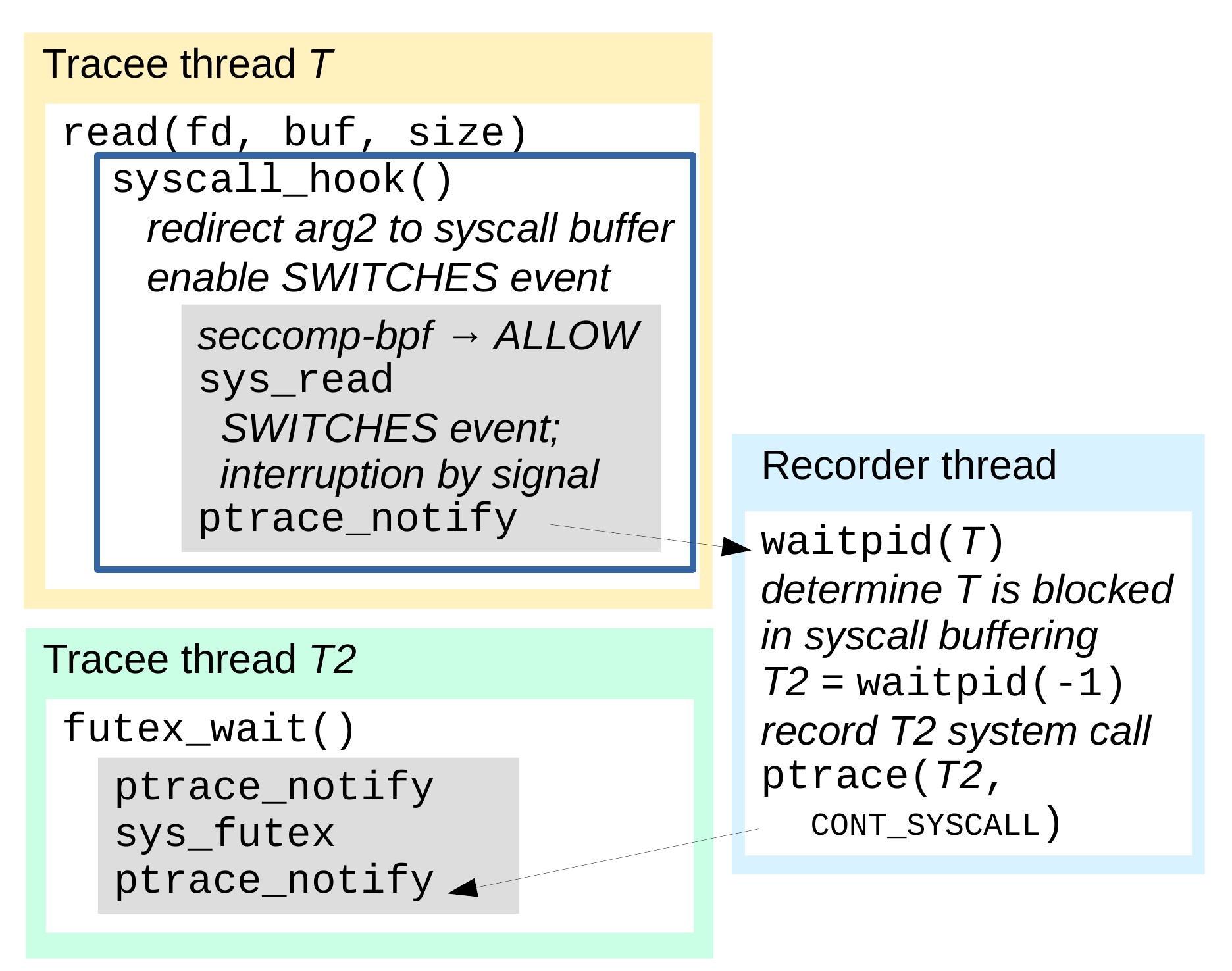}
\caption{Recording a blocking system call}
\label{syscallbuf-blocked-diagram}
\end{figure}

Figure \ref{syscallbuf-blocked-diagram} illustrates recording a blocking {\tt read} system call with system-call interception. The kernel deschedules the thread, triggering a {\tt perf} event which sends a signal to the thread, \emph{rescheduling} it, interrupting the system call, and sending a \ptrace{} notification to the recorder. The recorder does bookkeeping to note that an intercepted system call was interrupted in thread {\it T}, then checks whether any tracee threads in blocking system calls have progressed to a system-call exit and generated a \ptrace{} notification. In this example {\it T2} has completed a (not intercepted) blocking {\tt futex} system call, so we resume executing {\it T2}.

Meanwhile the {\tt read} system call in {\it T} is restarted and treated much like a regular non-intercepted system call. Its
parameters are rewritten to use scratch buffers. {\it T} is resumed using {\tt PTRACE\_SYSCALL} so that the thread will stop when the system call exits. After the system call exits, {\it T} will eventually be rescheduled. At that point
the system-call exit will be recorded. Because the system call was recorded normally, it should not also be recorded in the
interception library's trace buffer. Therefore \system{} sets an ``abort commit'' flag in the interception
library's memory before resuming.

\subsection{Signal Handling}

If interception code was interrupted by a context switch or asynchronous signal partway through recording
the result of a system call, that would be disastrous. When {\tt ptrace} reports a signal arriving at a thread
in interception code, \system{} stashes the signal away and sets a flag
to indicate that the interception code must perform a traced ``dummy'' system call just before returning to application
code, then resumes the tracee without delivering the signal. When the interception code performs a traced
system call (the ``dummy'' system call, or some other system
call for which we don't have a fast self-recording path), \system{} arranges to drain stashed
signals before the traced system call is allowed to proceed (by setting a breakpoint at the traced
system call entry point while there are stashed signals).

A difficult edge case involves untraced {\tt sigprocmask} system calls.
The kernel may try to deliver a signal to a thread in the interception library
just before {\tt sigprocmask} blocks the signal. If \system{}
were to stash the signal and try to inject it after the {\tt sigprocmask}, the signal might not be deliverable.
Therefore, while there are stashed signals, \system{} sets a breakpoint on the untraced system call
instruction. If a signal arrived before an untraced {\tt sigprocmask}, the breakpoint is hit and
\system{} emulates the {\tt sigprocmask} returning {\tt EAGAIN}, causing the interception code
to automatically retry it using a traced system call, which gives \system{} an opportunity to
inject all stashed signals (as noted above) and always succeeds.

Another difficult edge case involves multiple signals.
Normal signal delivery can automatically block future signals, sometimes permanently 
(e.g.\ using the {\tt sigaction} {\tt sa\_mask} feature). For example suppose signals $S_1$
and then $S_2$ are sent to a process with two threads $T_1$ and $T_2$, and the handler for $S_1$
is configured to block $S_2$ and just exits the thread.
Then suppose under \system{} the kernel delivers $S_1$ to $T_1$ while $T_1$ is in
interception code; \system{} stashes $S_1$ and resumes $T_1$. If the kernel were to then
deliver $S_2$ to $T_1$ we would have a problem, because $S_2$ will never be handled by
any thread, whereas during non-recorded execution the kernel would have delivered it to $T_2$ instead.
There is not enough information available from user-space to determine where the kernel
would have delivered $S_2$ if it had been blocked in $T_1$ (in particular we can't tell the difference
between a signal dispatched to a specific thread and a signal dispatched to a process).

Therefore it is important to ensure the kernel does not deliver a signal to a recorded thread
unless that thread will definitely be able to handle the signal. While there is a stashed signal
for a thread, we block any more signals from being delivered to the thread. This means there
can never be more than one stashed signal for a thread.

\subsection{Interrupted System Calls}

An untraced system call can be interrupted by a signal causing a signal handler to run, after which
the untraced system call may be resumed. This is
similar to handling a blocked untraced system call as described above, and we handle it much the
same way. The main extra complication is that the signal handler can run application code while
interception library code is on the stack. To prevent reentry into the interception library code,
we set a thread-local ``locked'' flag to indicate that any intercepted system calls should avoid
trying to take the untraced fast path until the signal handler has returned.

\subsection{Thread-Local Storage}

The interception library code needs access to thread-local data. Unfortunately in some rare
cases the standard Linux thread-local data machinery is not set up correctly, e.g.\ when
certain applications (Chromium!) use the raw {\tt clone} system call to create a thread. To work around this,
\system{} allocates a special ``thread locals'' memory page in each process at a fixed address, and
on every context switch copies out the old thread's values and copies in the new thread's values for
that page.

This approach also simplifies handling of {\tt fork}. Unlike regular Linux thread-local storage,
we make our thread-locals start off initialized to zero after a {\tt fork}. Thus when calling into
the interception library after {\tt fork}, it can reinitialize itself.

\subsection{Stack Handling}

On x86-64, leaf functions are allowed to use a 128-byte ``red zone'' of memory
below the stack pointer to avoid the cost of adjusting the stack pointer. The interception
library needs to avoid modifying that memory. Using memory below the red zone is dangerous because
it might cause a spurious stack overflow. (Go programs with many small stacks are particularly
vulnerable to this.) Therefore the trampolines that enter the interception library also switch stacks
to a temporary stack allocated as part of the thread's scratch buffer.

In theory this stack switching could cause problems for application signal handlers that expect
to be called on the regular stack but instead see our switched stack. In practice this has not
been a problem yet. If it was a problem, we would probably have to drastically modify our
strategy for application signal handling so that we completely unwind the interception library
stack frames and run the application signal handler as if the original system call instruction had been
interrupted.

\subsection{Handling Replay}

Conceptually, during recording we need to copy system call output buffers to a trace buffer, and during replay
we need to copy results from the trace buffer to system call output buffers. This is a problem
because the interception library is part of the recording and replay and therefore should
execute the same code in both cases. (Previous work with user-level system call interception \cite{Chastain1999, Saito2005, Geel2006, Mashtizadeh2017} avoided these problems by having less strict goals for replay fidelity.)

For this reason (and to avoid races of the sort discussed in Section \ref{syscalls}),
the interception library redirects system call outputs to write directly to the trace buffer.
After the system call completes, the interception library copies the output data from the trace buffer to
the original output buffer(s). During replay the untraced system call instruction is replaced with a no-op, so the system call
does not occur; the results are already present in the trace buffer so the post-system-call copy from the
trace buffer to the output buffer(s) does what we need.

During recording, each untraced system call sets a result register and the interception library writes it to the
trace buffer. Replay must read the result register from the trace buffer instead. We use a conditional move instruction so that control flow is perfectly consistent
between recording and replay. The condition is loaded from an {\tt is\_replay} global variable, so the
register holding the condition is different over a very short span of instructions (and explicity cleared afterwards).

Handling ``in-out'' system call memory parameters is tricky. During
recording we copy the input buffer to the trace buffer, pass the system call a
pointer to the trace buffer, then copy the trace buffer contents back to the input
buffer. Performing that first copy during replay would overwrite the trace buffer values
holding the system call results, so during replay we turn that copy into a no-op using
a conditional move to set the source address copy to the destination address.

We could allow replay of the interception library to diverge further from its
recorded behavior, but that would have to be done very carefully. We'd have to ensure
the RCB count was identical along both paths, and that register values
were consistent whenever we exit the interception library or trap to \system{} within the
interception library. It's simplest to minimize the divergence.

Replay of the interception library could be simplified a little by taking a different
approach in which untraced system calls are turned into traced system calls during replay.
However, it is important for replay performance that replay of untraced system calls avoid context
switching to the \system{} supervisor process, and for our purposes replay performance is important.

\subsection{Optimizing Reads With Block Cloning}

When an input file is on the same filesystem as the recorded trace and the filesystem supports copy-on-write cloning of file blocks, for large block-aligned {\tt read}s the system call interception code clones the data to a per-thread ``cloned-data'' trace file, bypassing the normal system-call recording logic. This greatly reduces space and time overhead for file-read-intensive workloads; see the next section.

This optimization works by cloning the input blocks and then reading the input data from the original input file. This opens up a possible race: between the clone and the read, another process could overwrite the input file data, in which case the data read during replay would differ from the data read during recording, causing replay to fail. However, when a file read races with a write under Linux, the reader can receive an arbitrary mix of old and new data, so such behavior would almost certainly be a severe bug, and in practice such bugs do not seem to be common. The race could be avoided by reading from the cloned-data file instead of the original input file, but that performs very poorly because it defeats Linux's readahead optimizations (since the data in the cloned-data file is never available until just before it's needed).

\subsection{Retrospective}

In-process system-call interception looks reasonably simple in principle, but in practice it has been
a large maintenance burden. It has to work unchanged during recording and replay, it has to behave like
a regular system call as much as possible (e.g. being atomic with respect to signals), and it has to be protected
from various sorts of application misbehavior. It cannot be protected from malicious applications. On the
other hand, without it we could not meet our performance targets.

\section{Results} \label{results}

\subsection{Workloads}

Benchmarks were chosen to illuminate \system{}'s strengths and weaknesses, while also containing representatives of real-world usage. They were tuned to fit in system memory (to minimize the impact of I/O on test results), to run for about 30 seconds each (except for \emph{cp} where a 30s run time would require it to not fit in memory).

\emph{cp} duplicates a {\tt git} checkout of {\tt glibc} (revision 2d02fd07) using {\tt cp -a} (15200 files constituting 732MB of data, according to {\tt du -h}). {\tt cp} is single-threaded, making intensive use of synchronous reads and a variety of other filesystem-related system calls.

\emph{make} builds DynamoRio \cite{Bruening2012} (version 6.1.0) with {\tt make -j8} ({\tt -j8} omitted when restricting to a single core). This tests potentially-parallel execution of many short-lived processes.

\emph{octane} runs the Google Octane benchmark under the Mozilla Spidermonkey Javascript engine (Mercurial revision 9bd900888753). This illustrates performance on CPU-intensive code in a complex language runtime.

\emph{htmltest} runs the Mozilla Firefox HTML forms tests (Mercurial revision 9bd900888753). The harness is excluded from recording (using {\tt mach mochitest -f plain --debugger \system{} dom/html/test/forms}). This is an example from real-world usage. About 30\% of user-space CPU time is in the harness.

\emph{sambatest} runs a Samba (git revision 9ee4678b) UDP echo test via {\tt make test TESTS=samba4.echo.udp}. This is an example from real-world usage.

All tests run on a Dell XPS15 laptop with a quad-core Intel Skylake CPU (8 SMT threads), 16GB RAM and a 512GB SSD using Btrfs in Fedora Core 23 Linux.

\subsection{Overhead}

\begin{table*}[t]
  \centering
  \begin{tabular}{lrrrrrrrr}
Workload & \shortstack[r]{Baseline \\ duration} & Record & Replay & \shortstack[r]{Single \\ core} & \shortstack[r]{Record \\ no-intercept} & \shortstack[r]{Replay \\ no-intercept} & \shortstack[r]{Record \\ no-cloning} & \shortstack[r]{DynamoRio- \\ null} \\
\hline
cp & 1.04s & 1.49$\times$ & 0.72$\times$ & 0.98$\times$ & 24.53$\times$ & 15.39$\times$ & 3.68$\times$ & 1.24$\times$ \\
make & 20.99s & 7.85$\times$ & 11.93$\times$ & 3.36$\times$ & 11.32$\times$ & 14.36$\times$ & 7.84$\times$ & 10.97$\times$ \\
octane & 32.73s & 1.79$\times$ & 1.56$\times$ & 1.36$\times$ & 2.65$\times$ & 2.43$\times$ & 1.80$\times$ & crash \\
htmltest & 23.74s & 1.49$\times$ & 1.01$\times$ & 1.07$\times$ & 4.66$\times$ & 3.43$\times$ & 1.50$\times$ & 14.03$\times$ \\
sambatest & 31.75s & 1.57$\times$ & 1.23$\times$ & 0.95$\times$ & 2.23$\times$ & 1.74$\times$ & 1.57$\times$ & 1.43$\times$ \\
  \end{tabular}
  \caption{Run-time overhead}
  \label{run-time-overhead-table}
\end{table*}

Table \ref{run-time-overhead-table} shows the wall-clock run time of various configurations, normalized to the run time of the baseline configuration. \emph{octane} is designed to run for a fixed length of time and report a score, so we report the ratio of the baseline score to the configuration-under-test score --- except for replay tests, where the reported score will necessarily be the same as the score during recording. For \emph{octane} replay tests we report the ratio of the baseline score to the recorded score, multiplied by the ratio of replay run time to recording run time.
Each test was run six times, discarding the first result and reporting the geometric mean of the other five results. Thus the results represent warm-cache performance.

``Single core'' reports the overhead of just restricting all threads to a single core using Linux {\tt taskset}.

``Record no-intercept'' and ``Replay no-intercept'' report overhead with in-process system-call interception disabled (which also disables block cloning). ``Record no-cloning'' reports overhead with just block cloning disabled.

``DynamoRio-null'' reports the overhead of running the tests under the DynamoRio \cite{Bruening2012} (version 6.1.0) ``null tool'', to estimate a lower bound for the overhead of using dynamic code instrumentation as an implementation technique. (DynamoRio is reported to be among the fastest dynamic code instrumentation engines.)

\subsection{Observations}

Overhead on \emph{make} is significantly higher than for the other workloads. Forcing \emph{make} onto a single core imposes major slowdown. Also, \emph{make} forks and execs 2430 processes, mostly short-lived. (The next most prolific workload is \emph{sambatest} with 89.) In-process system-call interception only starts working in a process once the interception library has been loaded, but at least 80 system calls are performed before that completes, so its effectiveness is limited for short-lived processes.

\begin{figure}[t]
\includegraphics[scale=0.4]{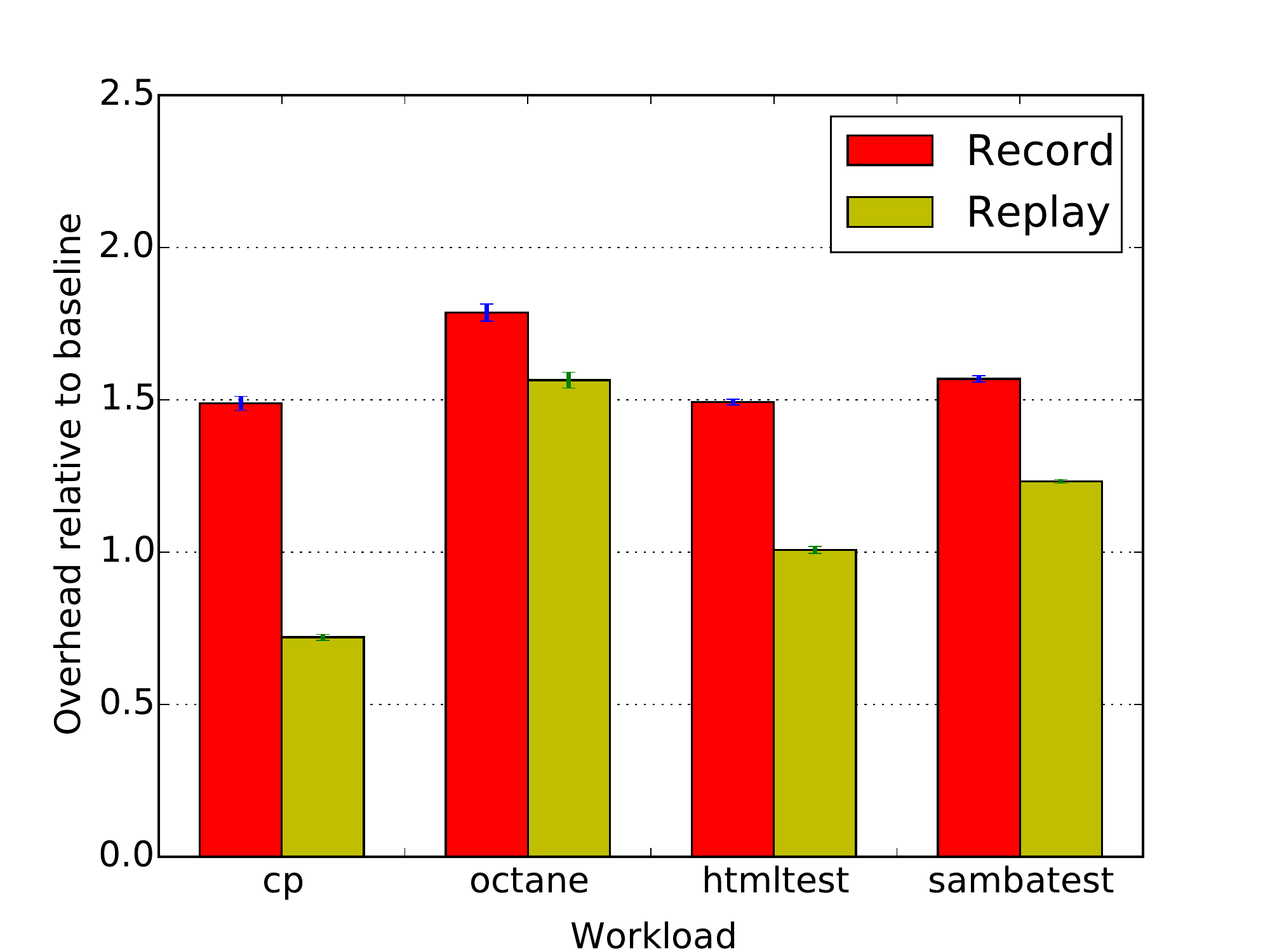}
\caption{Run-time overhead excluding \emph{make}}
\centering
\label{run-time-overhead-chart}
\end{figure}

Figure \ref{run-time-overhead-chart} shows the overall recording and replay overhead for workloads other than \emph{make}. Error bars in figures show 95\% confidence intervals; these results are highly stable across runs.

Excluding \emph{make}, \system{}'s recording slowdown is less than a factor of two. Excluding \emph{make}, \system{}'s replay overhead is lower than its recording overhead. Replay can even be faster than normal execution, in \emph{cp} because system calls do less work. For interactive applications, not represented here, replay can take much less time than the original execution because idle periods are eliminated.

\emph{octane} is the only workload here other than \emph{make} making significant use of multiple cores, and this accounts for the majority of \system{}'s overhead on \emph{octane}.

\begin{figure}[t]
\includegraphics[scale=0.4]{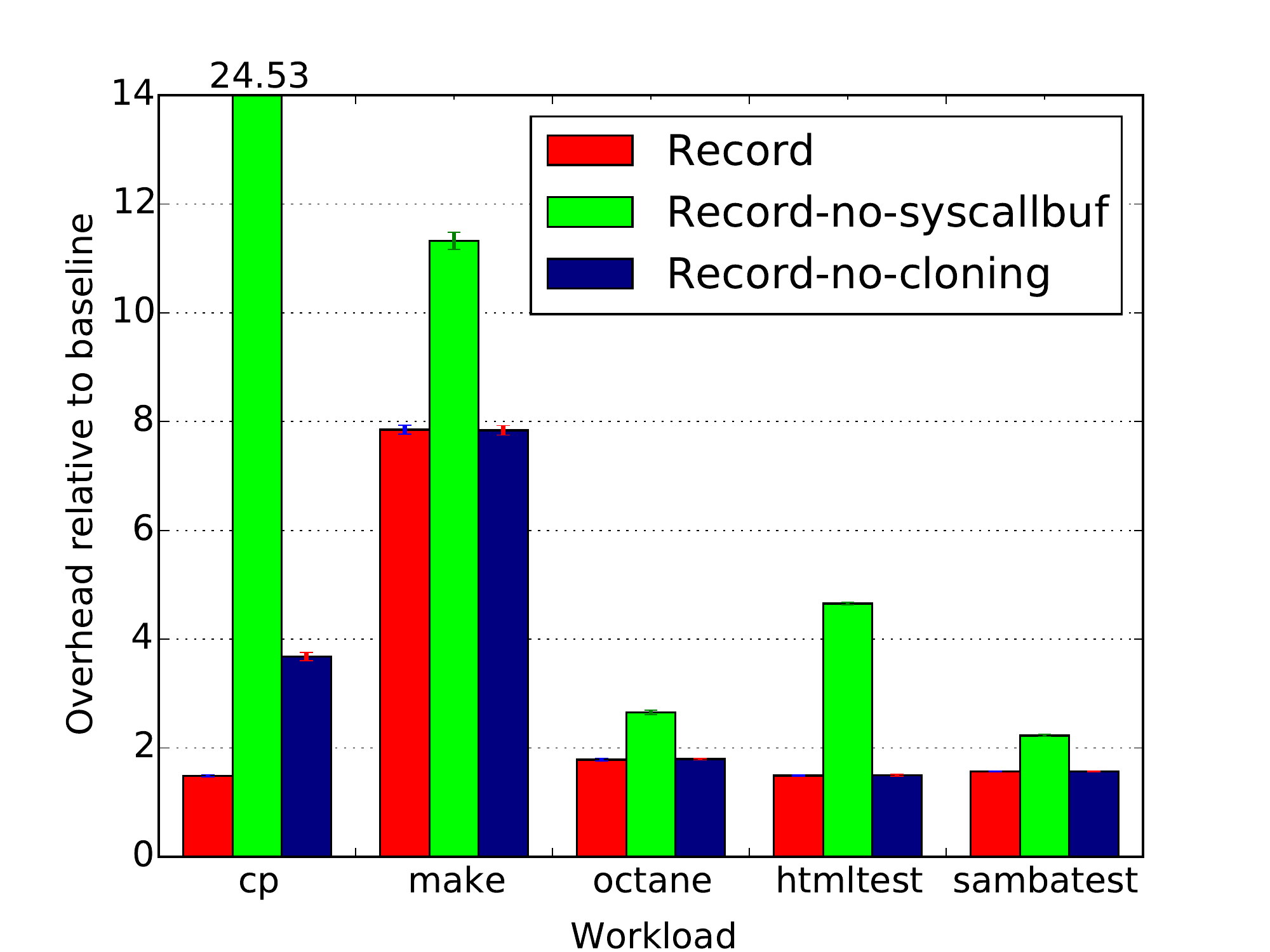}
\caption{Impact of optimizations}
\centering
\label{optimizations-chart}
\end{figure}

Figure \ref{optimizations-chart} shows the impact of system-call interception and blocking cloning on recording. The system-call interception optimization produces a large reduction in recording (and replay) overhead. Cloning file data blocks is a major improvement for \emph{cp} recording but has essentially no effect on the other workloads.

\begin{figure}[t]
\includegraphics[scale=0.4]{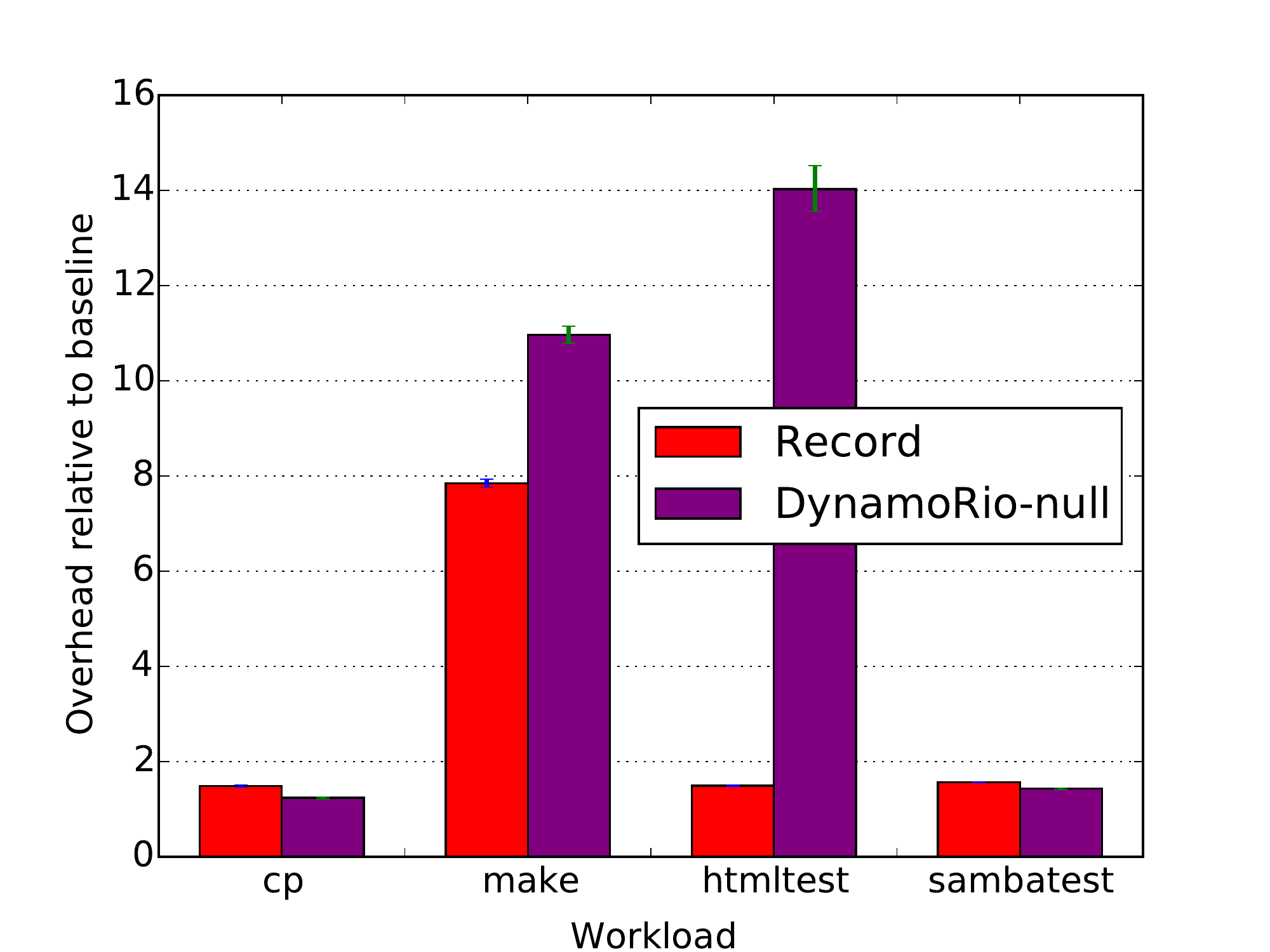}
\caption{Comparison with DynamoRio-null}
\centering
\label{dynamorio-chart}
\end{figure}

Figure \ref{dynamorio-chart} compares \system{} recording overhead with DynamoRio's ``null tool'', which runs all code through the DynamoRio instrumentation engine but does not modify the code beyond whatever is necessary to maintain supervised execution; this represents a minimal-overhead code instrumentation configuration. DynamoRio crashed on \emph{octane}
\footnote{We reported DynamoRio's crash on our ``octane'' workload to the developers at \url{https://github.com/DynamoRIO/dynamorio/issues/1930}.}. \emph{cp} executes very little user-space code and DynamoRio's overhead is low on that workload. On \emph{make} and \emph{sambatest} DynamoRio overhead is similar to \system{} recording, even though on \emph{make} DynamoRio can utilize multiple cores. On \emph{htmltest} DynamoRio's overhead is very high, possibly because that test runs a lot of Javascript with dynamically generated and modified machine code. Implementing record-and-replay on top of dynamic instrumentation would incur significant additional overhead, so we would expect the resulting system to have significantly higher overhead than \system{}.

\subsection{Storage Space Usage} \label{space-results}

\system{} traces contain three kinds of data: cloned (or hard-linked) files used for memory-map operations, cloned file blocks, and all other trace data, especially event metadata and the results of general system calls.

Memory-mapped files are mostly the executables and libraries loaded by tracees. While the original files are not changed or removed, which is usually true in practice, their clones take no additional space and require no data writes. \system{} makes no attempt to consolidate duplicate file clones, so most traces contain many duplicates and reporting meaningful space usage for these files is both difficult and unimportant in practice. The same is true for cloned file blocks.

\begin{table}[t]
  \centering
  \begin{tabular}{lrrr}
Workload & \shortstack[r]{Compressed \\ MB/s} & \shortstack[r]{{\tt deflate} \\ ratio} & \shortstack[r]{Cloned blocks \\ MB/s} \\
\hline
cp & 19.03 & 4.87$\times$ & 586.14 \\
make & 15.82 & 8.32$\times$ & 5.50 \\
octane & 0.08 & 8.33$\times$ & 0.00 \\
htmltest & 0.79 & 5.94$\times$ & 0.00 \\
sambatest & 6.85 & 21.87$\times$ & 0.00 \\
  \end{tabular}
  \caption{Storage space usage}
  \label{space-usage-table}
\end{table}

Table \ref{space-usage-table} shows the storage usage of each workload, in MB/s, for general trace data and cloned file blocks. We compute the geometric mean of the data usage for each trace and divide by the run-time of the workload baseline configuration. Space consumption shows very little variation between runs.

Different workloads have highly varying space consumption rates, but several MB/s is easy for modern systems to handle. In real-world usage, trace storage has not been a concern.

\subsection{Memory Usage}

\begin{table}[t]
  \centering
  \begin{tabular}{lrrrr}
Workload & Baseline & Record & Replay & Single core \\
\hline
cp & 0.51 & 34.54 & 9.11 & 0.51 \\
make & 510.51 & 327.19 & 314.16 & 288.65 \\
octane & 419.47 & 610.48 & 588.01 & 392.95 \\
htmltest & 690.81 & 692.06 & 324.71 & 689.75 \\
sambatest & 298.68 & 400.49 & 428.79 & 303.03 \\
  \end{tabular}
  \caption{Memory usage (peak PSS MB)}
  \label{mem-usage-table}
\end{table}

\begin{figure}[t]
\centering
\includegraphics[scale=0.4]{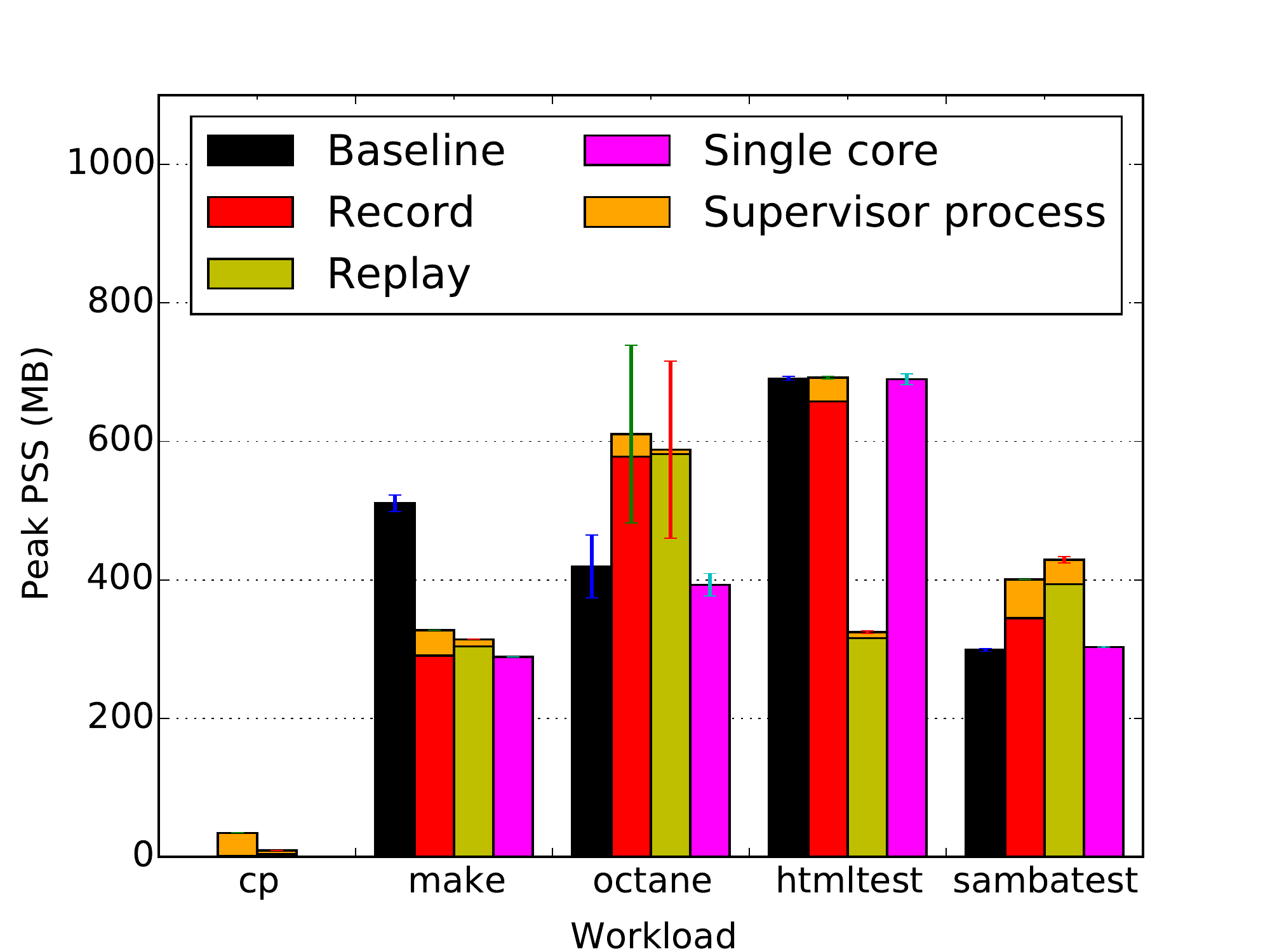}
\caption{Memory usage}
\label{mem-usage-chart}
\end{figure}

Table \ref{mem-usage-table} shows the memory usage of each workload. Every 10ms we sum the proportional-set-size (``PSS'') values of all workload processes (including \system{} if running); we determine the peak values for each run and take their geometric mean. In Linux, each page of memory mapped into a process's address space contributes $1/n$ pages to that process's PSS, where $n$ is the number of processes mapping the page; thus it is meaningful to sum PSS values over processes which share memory. The same data are shown in Figure \ref{mem-usage-chart}. In the figure, the fraction of PSS used by the \system{} process is shown in orange. Memory usage data was gathered in separate runs from the timing data shown above, to ensure the overhead of gathering memory statistics did not impact those results.

Given these experiments ran on an otherwise unloaded machine with 16GB RAM and all data fits in cache, none of these workloads experienced any memory pressure. \emph{cp} uses almost no memory. In \emph{make}, just running on a single core reduces peak PSS significantly because not as many processes run simultaneously. In \emph{octane} memory usage is volatile (highly sensitive to small changes in GC behavior) but recording significantly increases application memory usage; recording also increases application memory usage a small amount in \emph{sambatest} but slightly decreases it in \emph{htmltest}. (We expect a small increase in application memory usage due to system-call interception and scratch buffers.) These effects are difficult to explain due to the complexity of the applications, but could be due to changes in timing and/or effects on application or operating system memory management heuristics.

Replay memory usage is similar to recording except in \emph{htmltest}, where it's dramatically lower because we're not replaying the test harness.

\system{}'s memory overhead is not an issue in practice.

\section{Hardware/Software Design Constraints} \label{constraints}

We summarize the hardware and software features \system{} depends on, for system designers who may be interested in supporting \system{}-like record-and-replay.

\subsection{Hardware}

As discussed in Section \ref{counters}, \system{} requires a ``determinstic'' hardware performance
counter to measure application progress. The ideal performance counter for
our purposes would count the exact number of instructions retired as observed in user-space
(e.g., counting an interrupted-and-restarted instruction once).
Virtual machines should support reliable performance-counter virtualization.
Currently \system{} works under KVM and VMware, but VMware's
\emph{VM exit clustering} optimization \cite{Agesen2012},
as implemented, breaks the determinism of the RCB
counter and must be manually disabled.

Some x86 CPU instructions are nondeterministic. Section \ref{instructions} discusses our
current workarounds for this. Exposing hardware and OS support for trapping {\tt CPUID} is
important for long-term control over these instructions.

We would like to support record-and-replay of programs using hardware transactional
memory ({\tt XBEGIN}/{\tt XEND}). It would suffice if hardware
and the OS could be configured to raise a signal on any failed transaction.

Trapping on all other nondetermnistic instructions (e.g.\ {\tt RDRAND}) would
be useful.

Porting \system{} to ARM failed because all ARM atomic memory operations
use the ``load-linked/store-conditional'' approach, which is inherently nondeterminstic.
The conditional store can fail because of non-user-space-observable activity, e.g.\ hardware interrupts, so counts of retired instructions or conditional branches for
code performing atomic memory operations are nondeterminstic. These
operations are inlined into very many code locations, so it appears
patching them is not feasible except via pervasive code instrumentation or compiler changes.
On x86(-64), atomic operations (e.g.\ compare-and-swap) are deterministic in terms of
user-space state, so there is no such problem.

\subsection{Software}

As noted in Section \ref{shmem}, \system{} depends on configuring applications to avoid sharing memory with non-recorded processes.

We described how \system{} performance depends on modern Linux features: {\tt seccomp-bpf} to selectively trap system calls, {\tt PERF\_COUNT\_SW\_CONTEXT\_SWITCHES} performance events to handling blocking system calls, and copy-on-write file and block cloning APIs to reduce I/O overhead.

Efficient record-and-replay depends on clearly identifying a boundary within which code is replayed deterministically, and recording and replaying the timing and contents of all inputs into that boundary. In \system{}, that boundary is mostly the interface between the kernel and user-space. This suits Linux: most of the Linux user/kernel interface is stable across OS versions, relatively simple and well-documented, and it's easy to count hardware performance events occurring within the boundary (i.e.\ all user-space events for a specific process). This is less true in other operating systems. For example, in Windows, the user/kernel interface is not publicly documented, and is apparently more complex and less stable than in Linux. Implementing and maintaining the \system{} approach for Windows would be considerably more challenging than for Linux, at least for anyone other than the OS vendor.

\system{} relies on models that specify for each system call, how to efficiently determine from the before-and-after user-space state which memory locations were modified by the system call. The smaller the number of different
system-call interfaces (e.g.\ using APIs based on {\tt read} and {\tt write} through file descriptors instead of
arbitrary {\tt ioctl}s), the less work that is.

\section{Lessons Learned} \label{reflection}

\subsection{Deployability and Usage}

We set out to build a record-and-replay system optimized for deployability, and we seem to have
achieved that. Some Linux distributions package \system{}, and it's easy to download or build
for those that don't. On many distributions users must tweak the {\tt perf\_events\_paranoid} kernel
setting to enable non-root access to {\tt PERF\_COUNT\_SW\_CONTEXT\_SWITCHES}, but otherwise
\system{} ``just works'' for unprivileged users running most applications in non-virtualized
environments. This low barrier to entry is extremely important, since encouraging users to use
new tools is difficult under any circumstances.

We try hard to avoid requiring manual configuration. For example, \system{} automatically
detects kernel and virtualization bugs affecting it by running quick tests on every
launch; if found, it prints warnings with advice about how to
fix the environment (e.g.\ by upgrading software components) but automatically activates workarounds
for the bugs.

Our goal was always to work very well for a significant set of users. We chose good performance
for low-parallelism applications (and high slowdown for highly parallel applications) over
mediocre performance for all applications because we felt the former would please more users,
especially at Mozilla. It still appears we made the right choice.

\system{} has been mostly feature-complete for at least two years. Some of the authors have
been developing new tools on top of \system{}, but for the core open source project we have been
focusing on reliability and maintainance to ensure \system{} keeps working well as hardware,
Linux and applications evolve. Avoiding scope creep has helped keep the complexity of \system{}
under control, and therefore contributed to reliability. Reliability matters
because it's very important to users that
\system{} not fall over during a long debugging session. It's a lot more important
that \system{} work 100\% of the time on 90\% of applications than 90\% of the time on 100\% of
applications; the former pleases 90\% of users, but the latter pleases no-one.

Space and time overhead of \system{}, at least for low-parallelism workloads, seems to be
mostly a non-issue for our users. Reported performance problems tend to be due to system calls
not having a fast-path in the interception library, or just silly \system{} bugs,
though sometimes denying direct access to the GPU
causes performance problems. For many users the overhead of \system{} is less than the difference
between an optimized build and a debug build of their application, and therefore not very important.

Some record-and-replay techniques \cite{Altekar2009} reduce recording overhead in exchange for slower
replay, but for our users replay performance is very important --- often more important than recording
performance (e.g.\ when debugging an easily-reproduced bug, recording executes code once but replay
with reverse-execution may execute it many times.)

Some record-and-replay applications need to frequently create and resume checkpoints of replayed
execution state (e.g.\ to simulate reverse execution). Making checkpoints cheap in time and memory
is an often-overlooked design constraint. Anecdotally, creating and restoring checkpoints in VMWare's
record-and-replay debugger was rather slow, perhaps because the entire OS state must be checkpointed.
\system{}'s process-level design made it easy for us
to implement cheap checkpoints: we use {\tt fork} to copy address spaces and delay creating the
non-main threads until the checkpoint is resumed (and most checkpoints are
never resumed). {\tt fork} is (mostly) copy-on-write and is very well optimized on Linux, so
creating a checkpoint typically takes less than ten milliseconds. Although checkpointing
and restoring a general Linux process from user-space is rather complicated \cite{CRIU} because a lot
of kernel state may need to be saved and reestablished, checkpointing and restoring
\emph{for replay only} is relatively simple, because only a minimal amount of kernel state is
reproduced during replay.

\subsection{Engineering \system{}}

\system{} is complex and relies on poorly documented hardware and OS features. Over time we've
encountered many hardware, kernel and hypervisor bugs affecting \system{} and had to work
around them (while doing our best to get the underlying bugs fixed, too). Those bugs, and
many bugs in \system{} itself, are often difficult to diagnose and fix. In particular, small bugs
can cause small changes in replay state that don't cause observable divergence until
long after the root cause. To some extent
\system{} makes debugging easier for many at the cost of very difficult debugging for its
developers!

Therefore it is important to detect issues as soon after the root cause as possible.
\system{} has a high density of consistency assertions, most of which are checked even in release builds,
and many of which produce custom diagnostic messages for easier understanding in the field.
For example, unsupported system calls
or ioctls produce a message clearly identifying the problem and the offending system call. In general we try to
make our invariants as strict as possible, even stricter than may seem necessary, the better to
catch bugs and unexpected situations.

\system{} has a lot of logging code, which is built into debug and release builds so that end
users can enable it just by setting an environment variable. This makes it easier to get meaningful
reports from users, and we have been able to diagnose many bugs just by asking users to collect the
right logs.

Being free software has helped. A small community has developed around \system{} and some users have been
able to diagnose and fix the bugs they've found, particularly expanding system call coverage.

Inspecting the state of tracees during recording and replay is very important for debugging \system{}.
Therefore \system{} offers an ``emergency debugger'' feature, which can be manually triggered or
triggered automatically when an error occurs. The emergency debugger activates a {\tt gdb} server
(the same code \system{} uses to provide replay debugging) to which {\tt gdb} can attach to inspect
the register and memory state of the tracee, with full support for symbols, source mapping, etc.

To track down subtle errors in memory state during replay, \system{} supports taking checksums
and full checkpoints of memory at selected points during recording and comparing them with the replay.
We also support single-stepping tracees during recording and replay with logging of all register states,
to help narrow down divergences (though that can cause bugs to disappear).

To understand obscure details of Linux kernel behavior and bugs, we frequently have had to resort to
reading kernel source code. We have also used the kernel's {\tt ftrace} \cite{ftrace} feature several times.
Implementing something like \system{} on a system without access to full OS source code would be daunting.

\system{} has a fairly good suite of automated tests that run on every commit.
As of May 15, 2017, \system{} comprises 75028 lines of C and C++ code, of which 21403 lines belong
to the test suite. We have spent a lot of effort tracking down intermittent failures in the
test suite; this usually reveals bugs in the tests, but sometimes has revealed bugs in
\system{} or the kernel that would have been difficult to diagnose in the field. Since
users can easily run the tests, that helps us to identify issues in a user's environment
causing problems for \system{}.

\system{} is an excellent debugging tool, so we would like to be able to debug \system{} using \system{} itself. This is hard
because \system{} uses kernel features like \ptrace{} very intensively and supporting those features
well in \system{} is difficult. Furthermore, debugging \system{}
recording using \system{} usually wouldn't make sense because if recording workload $X$ doesn't work,
recording workload ``\system{} + $X$'' is even less likely to work. However, \system{}
replay is simpler than recording at the kernel level (since most tracee system calls don't run),
and we support \system{} recording and replaying of \system{} replay. This has been useful, although
keeping the multiple levels of \system{} straight in one's head can be confusing!

Anecdotally we have heard that \system{} has been used to help debug QEMU-based virtual machine
record and replay \cite{Dolan-Gavitt2015}. In turn, virtual-machine-based record and replay with a good
debugging experience could make debugging \system{} a lot easier. Because these approaches are quite
different, they could be very helpful for debugging each other.

\section{Related Work}

\subsection{Whole-System Replay}

ReVirt \cite{Dunlap2002} was an early project that recorded and replayed the execution of an entire virtual machine. VMware \cite{Malyugin2007} used the same approach to support record-and-replay debugging in VMware Workstation, for a time, but discontinued the feature. The full-system simulator Simics supports reverse-execution debugging via deterministic reexecution \cite{Engblom2010}. There have been efforts to add some record-and-replay support to QEMU \cite{Dolan-Gavitt2015,Dovgalyuk2012,Srinivasan2011} and Xen \cite{Dunlap2008, Burtsev2016}. Whole-system record-and-replay can be useful, but it is often inconvenient to hoist the application into a virtual machine. Many applications of record-and-replay require cheap checkpointing, and checkpointing a VM image is generally more expensive than checkpointing one or a few processes.

\subsection{Replaying User-Space With Kernel Support}

Scribe \cite{Laadan2010}, dOS \cite{Bergan2010} and Arnold \cite{Devecsery2014} replay a process or group of processes by extending the OS kernel with record-and-replay functionality. Kernel changes make maintenance and deployment more difficult --- unless record-and-replay is integrated into the base OS. But adding invasive new features to the kernel has risks, so if record-and-replay can be well implemented outside the kernel, moving it into the kernel may not be desirable.

\subsection{Pure User-Space Replay}

Pure user-space record-and-replay systems have existed since at least MEC \cite{Chastain1999}, and later Jockey \cite{Saito2005} and liblog \cite{Geel2006}. Those systems did not handle asynchronous event timing and other OS features. PinPlay \cite{Patil2010}, iDNA \cite{Bhansali2006}, UndoDB \cite{UndoDB} and TotalView ReplayEngine \cite{Gottbrath2008} use code instrumentation to record and replay asynchronous event timing. Unlike UndoDB and \system{}, PinPlay and iDNA instrument all loads, thus supporting parallel recording in the presence of data races and avoiding having to compute the effects of system calls, but this gives them higher overhead than the other systems. Compared to the other systems that support asynchronous events, \system{} achieves lower overhead by avoiding code instrumentation.

\subsection{Higher-Level Replay} 

Record-and-replay features have been integrated into language-level virtual machines. DejaVu \cite{Choi2001} added record-and-replay capabilities to the Jalape\~{n}o Java VM. Microsoft IntelliTrace \cite{Microsoft2013} instruments CLR bytecode to record high-level events and the parameters and results of function calls; it does not produce a full replay. Systems such as Chronon \cite{Deva2010} for Java instrument bytecode to collect enough data to provide the appearance of replaying execution, without actually doing a replay. Dolos \cite{Burg2013} provides record-and-replay for JS applications in Webkit by recording and replaying nondeterministic inputs to the browser. R2 \cite{Guo2008} provides record-and-replay by instrumenting library interfaces; handling data races or asynchronous events requires user effort to isolate the nondeterminism. Such systems are all significantly narrower in scope than the ability to replay general user-space execution.

\subsection{Parallel Replay}

Recording application threads running concurrently on multiple cores, with the possibility of data races, with low overhead, is extremely challenging. PinPlay \cite{Patil2010} and iDNA/Nirvana \cite{Bhansali2006} instrument shared-memory loads and report high overhead. SMP-ReVirt \cite{Dunlap2008} tracks page ownership using hardware page protection and reports high overhead on benchmarks with a lot of sharing. DoublePlay \cite{Veeraraghavan2011} runs two instances of the application and thus has high overhead when the application alone could saturate available cores. ODR \cite{Altekar2009} has low recording overhead but replay can be extremely expensive and is not guaranteed to reproduce the same program states. Castor \cite{Mashtizadeh2017} instruments synchronization code by modifying compilers and runtime systems, which creates barriers to easy deployment, and cannot replay reliably in the presence of data races.

The best hope for general, low-overhead parallel recording seems to be hardware support. Projects such as FDR \cite{Xu2003}, BugNet \cite{Narayanasamy2005}, Rerun \cite{Hower2008}, DeLorean \cite{Montesinos2008} and QuickRec \cite{Pokam2013} have explored low-overhead parallel recording hardware.

\section{Future Work}

Probably the main issue holding back \system{} deployment currently is the lack of support for
virtualized hardware performance counters in the major cloud providers. This is not really a technical
issue; those providers need to be convinced that the benefits of virtualizing counters (or at least
those used by \system{}) outweigh the costs and risks. Expanding the usage of \system{} is part of
our effort to address this problem.

Some use-cases for record-and-replay involve transporting recordings from one machine to
another, which \system{} does not really support yet. Once {\tt CPUID} faulting support is widely
available, that will be easier to implement.

\system{} perturbs execution, especially by forcing all threads onto a single core, and therefore can fail to reproduce bugs that manifest outside \system{}. We have addressed this problem by introducing a ``chaos mode'' that intelligently adds randomness to scheduling decisions, enabling us to reproduce many more bugs, but that work is beyond the scope of this paper. There are many more opportunities to enhance the recorder to find more bugs.

Putting record-and-replay support in the kernel would have performance benefits, e.g.\ reducing the cost of recording context switches. We may be able to find reusable primitives that can be added to kernels to improve the performance of user-space record-and-replay while being less invasive than a full kernel implementation. It would be particularly interesting to find small kernel extensions that could eliminate the need for in-process system call interception. Some way to record the data passing through {\tt copy\_to\_user} would be a good start.

Recording multiple processes running in parallel on multiple cores seems feasible if they do not share memory --- or, if they share memory, techniques inspired by SMP-ReVirt \cite{Dunlap2008}, dthreads \cite{Liu2011} or Castor \cite{Mashtizadeh2017} may work for some workloads on existing hardware. We hope that popularizing practical use
of record-and-replay will help support the economic argument for adding hardware support for
data-race recording \cite{Pokam2013}.

The applications of record-and-replay are perhaps more interesting and important than the base technology. For example, one can perform high-overhead dynamic analysis during replay \cite{Devecsery2014, Dolan-Gavitt2015, Patil2010}, potentially parallelized over multiple segments of the execution. With \system{}'s no-instrumentation approach, one could collect performance data such as sampled stacks and performance counter values during recording, and correlate that data with rich analysis generated during replay (e.g.\ cache simulation). Always-on record-and-replay would make finding and fixing bugs in the field much easier. Demonstrating compelling applications for record-and-replay will build the case for building support into commodity hardware and software.

\section{Conclusions}

The current state of Linux on commodity x86 CPUs enables single-core user-space record-and-replay with low overhead, without pervasive code instrumentation --- but only just. This is fortuitous; we use software and hardware features for purposes they were not designed to serve. It is also a recent development; five years ago {\tt seccomp-bpf} and the Linux file cloning APIs did not exist, and commodity architectures with a deterministic hardware performance counter usable from user-space had only just appeared (Intel Westmere)\footnote{Performance counters have been usable for kernel-implemented replay \cite{Dunlap2002,Olszewski2009} for longer, because kernel code can observe and compensate for events such as interrupts and page faults.}. By identifying the utility of these features for record-and-replay, we hope that they will be supported by an increasingly broad range of future systems. By providing an open-source, easy-to-deploy, production-ready record-and-replay framework we hope to enable more compelling applications of this technology.

\bibliography{Master}
\bibliographystyle{abbrv} 

\end{document}